\documentclass[useAMS,usenatbib,twocolumn]{mnras}
\usepackage{graphicx}
\usepackage{subfigure}
\usepackage[fleqn]{amsmath}
\usepackage{url}
\usepackage{soul}
\usepackage{xcolor}

\renewcommand{\Vec}[1]{ \mbox{\boldmath$ #1 $} }

\newcommand{\mpt}{\mathrm{.}}
\newcommand{\mcm}{\mathrm{,}}

\newcommand{\grale}{{\sc grale}}

\renewcommand{\Vec}[1]{ \mbox{\boldmath$ #1 $} }

\title[Extrapolating the projected potential]{Extrapolating the projected potential of gravitational lens models: property-preserving degeneracies}
\author[J. Liesenborgs, D. Perera and L.L.R. Williams]{Jori Liesenborgs$^1$\thanks{Corresponding author: jori.liesenborgs@uhasselt.be},
			  Derek Perera$^{2}$ and
              Liliya L.R. Williams$^{2}$  \\
              $^1$ UHasselt -- Flanders Make, Expertisecentrum voor Digitale Media, Wetenschapspark 2, B-3590, Diepenbeek, Belgium \\
              $^2$ School of Physics and Astronomy, University of Minnesota, 116 Church Street SE, Minneapolis, MN 55455, USA
              }

\begin{document}

\date{} % TODO
	
\pagerange{\pageref{firstpage}--\pageref{lastpage}} \pubyear{2024}

\maketitle \label{firstpage} 
\begin{abstract} 
		While gravitational lens inversion holds great promise to reveal the
		structure of the light-deflecting mass distribution, both light and dark,
		the existence of various kinds of degeneracies implies that care must be 
		taken when interpreting the resulting lens models. This article illustrates
		how thinking in terms of the projected potential helps to gain
		insight into these matters. Additionally it is shown explicitly how, when starting
		from a discretised version of the projected potential of one particular 
		lens model, the technique of 
		quadratic programming can be used to create a multitude of 
		equivalent lens models that preserve all or a subset of lens properties. This
		method is applied to a number of scenarios, showing the lack of grasp
		on the mass outside the strong lensing region, revisiting
		mass redistribution in between images and applying this to a recent model of
		the SDSS~J1004+4112 cluster, as well as illustrating the generalised mass 
		sheet degeneracy and source-position transformation.
        In the case of J1004 we show that this mass redistribution did not 
        succeed at completely eliminating a dark mass clump recovered by \grale{} 
        near one of the quasar images.
\end{abstract}


\section{Introduction}

	Apart from causing beautiful observations, the light deflection caused by the
    gravitational lens effect holds the promise of providing insight into the distribution
    of the matter responsible for this deflection, as well as for probing parameters
    of the cosmological model. To make this possible, one typically needs to try to
    invert the lens effect, e.g. try to reconstruct a model for the gravitational lens
    that is compatible with the observations.

    Over the years, several techniques for doing so have been developed, differing in
    the kinds of observations they use as input as well as in how the matter distribution
    it tries to reconstruct, is modeled. This ranges from statistical analyses of small
    deformations of background galaxies, i.e. weak lensing data, to the use of multiple,
    possibly highly deformed images, also referred to as a strong lensing scenario.
    The cause of the deflection, the matter distribution of the lens itself, can be modelled
    by a relavitely small number of density profiles, typically aligned with the visible
    matter (e.g. {\sc LensTool}, \citet{2007NJPh....9..447J}), 
    by a large set of basis functions, intended to be capable of modelling a wide
    variety of distributions (e.g. {\sc PixeLens}, \citet{2004AJ....127.2604S}, \citet{2008ApJ...679...17C}),
    or even by both of these options, in a more hybrid approach (e.g. {\sc wslap+}, \citet{2014MNRAS.437.2642S}).
    Others still do not model the mass distribution directly, but instead model the lens'
    gravitational potential (e.g. {\sc relensing}, \citet{2023MNRAS.518.4494T}).

    Irrespective of the procedure and input data that are used, it is important to realize
    that a solution to the inversion problem is not uniquely determined, that there exist
    various kinds of degenerate solutions that are able to explain the observations equally
    well. Some of these are exact by nature, others differ in principle but only cause changes
    that still lie within the known observational uncertainties. Depending on how the mass
    distribution is modelled, these degeneracies can manifest themselves in different ways.
    It may even seem that there are no such degeneracies present, if the inversion technique 
    used does not provide the freedom needed to describe equivalent solutions. The existence
    of multiple, equally compatible solutions is fundamental however, so care must be taken
    when interpreting inversion results.

    In this article we illustrate how thinking about lens inversion not on the level of the
    mass distribution itself, but the potential causing it, can help gain additional
    insight into which properties can actually be constrained well. Assuming that one
    solution to the lens inversion problem is known, a tool is introduced
    that uses quadratic programming to search for lens models that are equally compatible
    with the observed data. 

    After briefly reiterating the gravitational lensing formalism in section \ref{sec:formalism},
    a toy model will be introduced in section \ref{sec:toymodel} to study an effect that is often
    encountered when performing lens inversions with our own method \grale, namely mass density
    peaks outside of the region covered by the multiple image systems. The idea behind the
    method and its practical implementation using quadratic programming is explained in sections
    \ref{sec:extrap} and \ref{sec:qp}. Application to the toy model for the outer density peaks, as well as
    revisiting known degeneracies using this method will be done in section \ref{sec:apps},
    ending the article with a final discussion in section \ref{sec:discussion}.

\section{Gravitational lensing formalism}\label{sec:formalism}

	Below, the formalism to describe gravitational lensing is briefly reviewed -- for a complete
    account the interested reader is referred to \citet{SchneiderBook}. In the usual approximation,
    the mass density of the gravitational lens itself is modelled as being two-dimensional,
    lying in the so-called lens plane. This mass density $\Sigma(\Vec{\theta})$, where $\Vec{\theta}$
    describes the viewing direction, causes light rays from source to observer to become deflected.
    The lens equation,
    \begin{equation}
        \Vec{\beta} = \Vec{\theta} - \frac{D_{\rm ds}}{D_{\rm s}} \Vec{\hat{\alpha}}(\Vec{\theta}) \mcm
		\label{eq:lenseqn}
    \end{equation}
    describes this mapping: when looking in direction $\Vec{\theta}$, one receives the light that
    one would receive from direction $\Vec{\beta}$ if the light deflection could be disabled somehow.
    The deflection angle $\Vec{\hat{\alpha}}(\Vec{\theta})$ describes the way the light ray changes
    direction due to the lens effect, and is determined by the projected mass density
    $\Sigma(\Vec{\theta})$ in its entirety. This deflection angle is rescaled by
    $D_{\rm ds}$ and $D_{\rm s}$, the angular diameter distances from gravitational lens to 
    source and from observer to source respectively. Similarly, the angular diameter distance from
    observer to lens will be denoted by $D_{\rm d}$. To ease notation, one often uses the rescaled
    deflection angle $\Vec{\alpha} = D_{\rm ds}/D_{\rm s}\; \Vec{\hat{\alpha}}$. The equation can be
    interpreted as describing how a two-dimensional source shape, lying in the so-called source plane
    and described by the $\Vec{\beta}$-space, is transformed into possibly multiple images lying
    in the image plane, described by the $\Vec{\theta}$-vectors.

    It can be shown that in this thin lens approximation, the deflection angle arises
    from a two-dimensional version of the gravitational potential, usually referred to as the lens
    potential or projected potential $\psi(\Vec{\theta})$:
    \begin{equation}
        \Vec{\alpha}(\Vec{\theta}) = \Vec{\nabla} \psi(\Vec{\theta}) \mpt
        \label{eq:gradpsi}
    \end{equation}

    This lens potential is also related to a scaled version $\kappa(\Vec{\theta})$ of the mass
    distribution
    \begin{equation}
        \kappa(\Vec{\theta}) = \frac{1}{2}\nabla^2 \psi(\Vec{\theta}) \mcm
        \label{eq:nablapsi}
    \end{equation}
    where $\kappa(\Vec{\theta}) = \Sigma(\Vec{\theta})/\Sigma_{\rm crit}$ is also called 
    the convergence, and $\Sigma_{\rm crit} = c^2 D_{\rm s}/4\pi G D_{\rm d} D_{\rm ds}$ is known
    as the critical density.

    If two image positions $\Vec{\theta}_i$ and $\Vec{\theta}_j$ correspond to the same source position $\Vec{\beta}$,
    there will be a time delay $\Delta t_{ij} = t(\Vec{\theta_i},\Vec{\beta}) - t(\Vec{\theta_j},\Vec{\beta})$
    between these images, which may be measurable for a source with intrinsic variability. Here,
    \begin{equation}
        t(\Vec{\theta},\Vec{\beta}) = \frac{1+z_d}{c} \frac{D_{\rm d} D_{\rm s}}{D_{\rm ds}} \left(\frac{1}{2}(\Vec{\theta}-\Vec{\beta})^2 - \psi(\Vec{\theta}) \right)
        \label{eq:timedelay}
    \end{equation}
    in which $z_d$ represents the redshift of the lens plane.

    The case where multiple images arise from a same source is called the strong lensing
    regime, but even when there is only a single image this can still be a somewhat deformed
    one. Further away from the bulk of the lensing mass, one then encounters the weak lensing
    regime. The deformations are described by the shear components
    \begin{equation}
        \gamma_1 = \frac{1}{2}\left(\frac{\partial^2 \psi}{\partial \theta_x^2} - \frac{\partial^2 \psi}{\partial \theta_y^2}\right)
        \textrm{, and }
        \gamma_2 = \frac{\partial^2\psi}{\partial \theta_x \partial \theta_y} \mpt
        \label{eq:gamma}
    \end{equation}
    Statistical analyses of deformed background galaxies cannot reveal these values directly
    unfortunately, only a combination of them with the convergence can be estimated at a
    point. This is then called the reduced shear $g_i = \gamma_i/(1-\kappa)$.

    For the remainder of the article, the focus will be on the strong lensing regime;
    the same ideas and procedures apply to the weak lensing regime as well.

\section{Outer density peaks}\label{sec:toymodel}

    \begin{figure}
        \centering
        \includegraphics[width=0.48\textwidth]{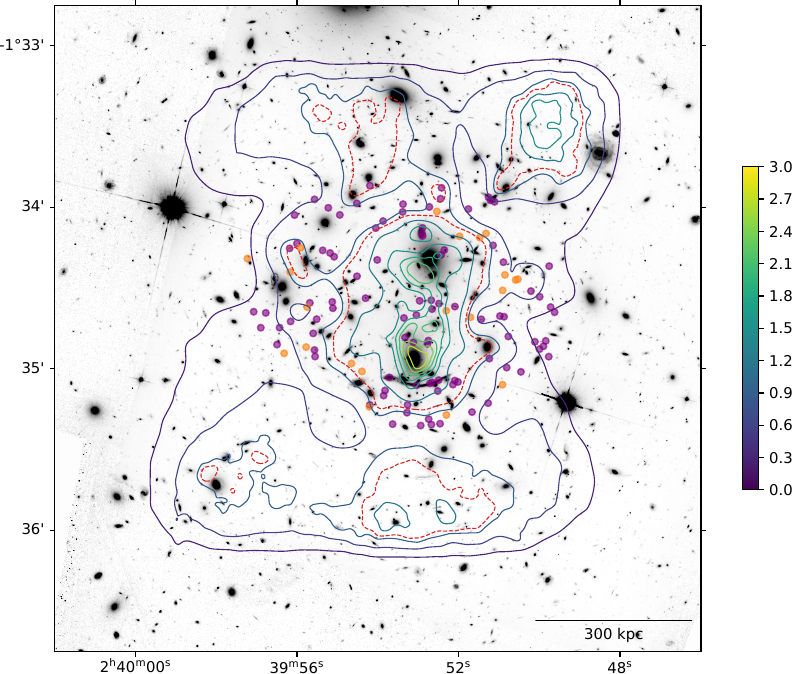}
        \caption{Reproduction of fig.~1 from \citet{2021MNRAS.506.6144G}, illustrating
        the effect that
        is often encountered when using the free-form \grale{} inversion method, when
        the optimization procedure places mass near the boundaries where the mass density
        is not well enough constrained by the more centrally located images.}
        \label{fig:externalmassexamples}
    \end{figure}

    In our free-form strong lens inversions using the \grale{} software
    \citep{Liesenborgs,2020MNRAS.494.3253L}, one has to specify
    the region where mass is to be recovered. Typically, this region should not exceed the
    boundaries set by the multiple image systems by too much, but the amount by which this
    is done in practice can vary somewhat: depending on the complexity of the lensing
    scenario one may need to make this region somewhat larger than a first estimate
    to be able to obtain a reconstruction that can explain the observed images adequately.

    In such cases the underlying optimization technique can place
    extra mass near the borders of the inversion region, where no multiple image systems
    enclose these structures. For a final solution several tens of optimization results
    are averaged, which does tend to smoothen, but not remove these structures. Fig.~\ref{fig:externalmassexamples}
    shows an example of this effect.
    It is commonly understood that as such mass density features are not enclosed by
    strongly lensed images, their precise location and shape should not carry too much
    weight. Instead, they can be interpreted as properties of the mass distribution that
    the inversion algorithm introduces to mimic external shear, but the precise origin
    of this shear cannot be constrained.

\begin{figure*}
        \centering
        \includegraphics[width=\textwidth]{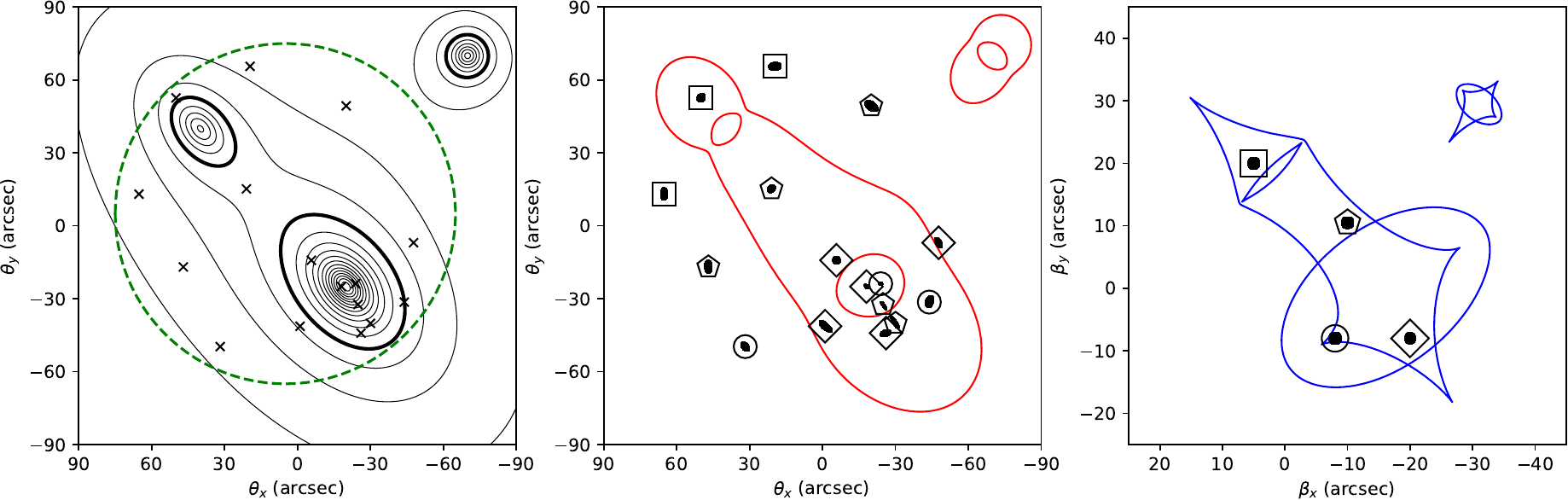}
        \caption{This is the toy model mentioned in section~\ref{sec:toymodel}.
        The left panel shows a mass distribution consisting of two centrally located
        non-singular isothermal ellipse (NSIE) profiles, where a mass peak
        has been added in the top-right corner of the region. The mass
        density is shown in units of $\kappa=\Sigma/\Sigma_{\rm cr}$ for
        a source at redshift $z=2.0$, the contours show levels of $\kappa$
        with intervals of $0.2$ where the $\kappa=1$ level is indicated
        by a thicker line. The dashed circle marks a rough 
        estimate of the strong lensing region, to be used in the procedure that shall attempt
        to erase the external features, i.e. the peak in the
        top-right corner. This mass distribution causes the 
        the sources at redshifts of 1.0 (diamond symbol), 1.5 (circle), 
        2.0 (square) and 2.5 (pentagon) from the right panel, to be lensed into the images that
        are shown in the center panel of this figure (the left panel marks these
        positions with crosses). Also shown in center and
        right panels are respectively the critical lines and caustics for a
        source at redshift $z=2.0$.}
        \label{fig:simpeak}
    \end{figure*}

    To illustrate the ill-constrained nature of the outer regions of a strong lensing scenario,
    the toy model from Fig.~\ref{fig:simpeak} will be used. The shape of the
    mass distribution is inspired by the Ares simulated cluster \citep{2017MNRAS.472.3177M},
    but has an extra mass peak in the top-right corner. The lens itself is
    located at a redshift of $z_d=0.5$ in a flat $\Lambda$CDM cosmological model with $H_0 = 70$ km s$^{-1}$ Mpc$^{-1}$
    and $\Omega_m = 0.3$, and causes the four circular sources
    shown in the right panel of the figure to be transformed
    into the images that can be seen in the center panel.

\section{Lens potential extrapolation}\label{sec:extrap}

    From the overview of the lensing formalism in section~\ref{sec:formalism},
    one can see that all properties can be derived from the lens potential
    $\psi(\Vec{\theta})$. More precisely, if two models have the same values
    of the lensing potential in the regions that contain all the images, from equation (\ref{eq:gradpsi}) 
	they will have the same deflection angles at those locations, and through the lens equation (\ref{eq:lenseqn}) will map 
    to the same positions in the source plane. With image locations, source
    location and lens potential values unchanged, equation (\ref{eq:timedelay}) implies
    that the time delays between images will be unchanged as well. Being
    very local properties derived from the lens potential, $\gamma_i$ values
    at the image locations will be the same (equation (\ref{eq:gamma})), as
    well as the $g_i = \gamma_i/(1-\kappa)$ values, since the convergence
    $\kappa$ stays unchanged by equation (\ref{eq:nablapsi}). 

    The central idea to resolve the presence of the mass peak in the toy model,
    is therefore to create a new model that has the same values of the lens 
    potential within the circular region indicated in the figure. As
    both models have the same\footnote{The models may actually differ by a constant value,
    which will have no observable effect.} potential values inside the circular
    region, all lensing properties will be conserved there. Outside the
    circle, the new model will have different values however:
    we shall start from the lens potential values
    inside the circle, and extrapolate these outward. The goal is to do this
    in such a way that said mass peak is less prominent or even erased.

    To be able to perform such calculations based on $\psi(\Vec{\theta})$ numerically,
    it will be necessary to approximate this continuous scalar field by a discrete
    grid of values $\psi_{ij}$. Within the modelling part of the \grale{} software,
    it is possible to specify a lens model based on such a grid. To calculate
    values of the first and second order derivatives, not only at the grid points
    themselves, but in between these points as well, the bicubic interpolation
    routines from the GNU Scientific Library\footnote{\url{https://www.gnu.org/software/gsl/}} (GSL)
    are used.

    \begin{figure*}
        \centering
        \includegraphics[width=\textwidth]{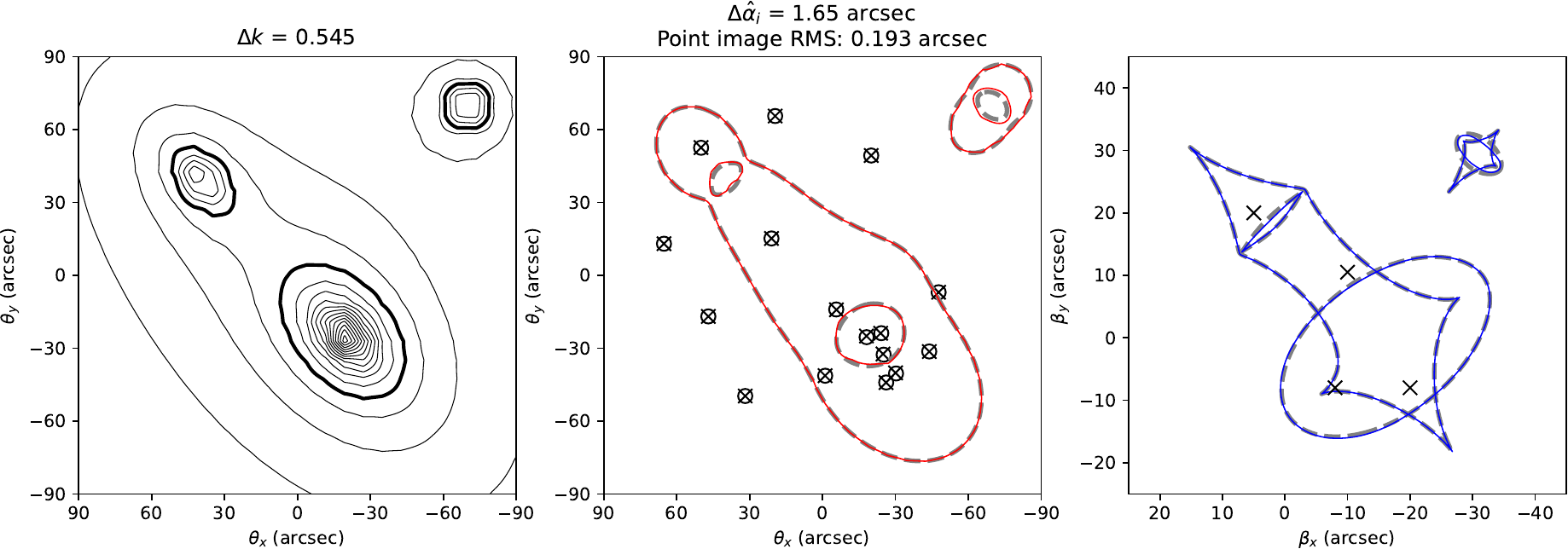}
        \caption{For the approximation of the lens from Fig.~\ref{fig:simpeak}
        based on lens potential values $\psi_{ij}$ on a $32\times 32$ grid,
        this figure again shows mass density, critical lines and caustics.
        For comparison, the dashed lines are those from Fig.~\ref{fig:simpeak}.
        Maximum differences in density $\Delta\kappa$ and deflection angle component $\Delta\hat{\alpha}_i$
        (calculated for the $512\times 512$ values that are plotted, so also in between
        the $\psi_{ij}$ grid points),
        as well as RMS difference in the image plane are also mentioned
        (the circles show the image positions according to this approximation,
        the crosses are those for the toy model lens itself).
        As for higher grid resolutions the visual differences from Fig.~\ref{fig:simpeak}
        become quickly unnoticeable, Table~\ref{tab:phiapproxprops} shows
        the maximum differences and RMS for these approximations.}
        \label{fig:phigridapprox}
    \end{figure*}

    In a first step, an approximation of the toy model is needed based on a grid of $\psi_{ij}$ values.
    To be able to calculate derivatives near the boundaries shown in the figure, a
    slightly larger region is used for this grid, in this case a $240\times240$ $\textrm{arcsec}^2$
    one. For a $32\times 32$ grid covering this region, Fig.~\ref{fig:phigridapprox} 
    shows results for this approximate lens model. In this case, differences are
    still noticeable, but as can also be seen in Table~\ref{tab:phiapproxprops}
    the approximation improves quite rapidly for higher $N_\psi \times N_\psi$ grid resolution.

    \begin{table}
       \begin{tabular}{l|lll}
            \hline
            $N_\psi$ & $\Delta\kappa$ & $\Delta\hat{\alpha}_i$ (arcsec) & Point image RMS (arcsec) \\
            \hline
            32 & 0.545 & 1.65 & 0.193 \\
            64 & 0.126 & 0.192 & 0.0174 \\
            128 & 0.0345 & 0.0235 & 0.00116 \\
            256 & 0.00879 & 0.00195 & $9.9\times 10^{-5}$ \\
            512 & 0.00131 & 0.000196 & $1.4 \times 10^{-5}$ \\
            \hline
        \end{tabular} 
        \caption{This table illustrates the properties mentioned in the
        caption of Fig.~\ref{fig:phigridapprox}
        for increasing $N_\psi \times N_\psi$ resolutions of the $\psi_{ij}$ grid.
        The $\Delta\kappa$ and $\Delta\hat{\alpha}_i$ numbers represent the maximum values
        of the differences.
        }
        \label{tab:phiapproxprops}
    \end{table}

	In what follows, the model based on $128\times 128$ values
	of $\psi_{ij}$ will be used as the starting point. This represents the true toy model
	with considerable accuracy, while yielding a manageable number of variables that will need to
	be handled in the extrapolation method described below.

\section{Quadratic programming solution}\label{sec:qp}

    For the simulated lensing scenario under consideration, the approach will be
    to keep the $\psi_{ij}$ values inside the circular region fixed, and to optimize
    for the other lens potential values. It shall turn out that this optimization
    can be formulated as a quadratic programming (QP) problem, for which several
    software packages exist to calculate the solution very efficiently.
    
    In this type of problem one looks for a vector $\Vec{x}$ of unknown values
    that mimimize the expression
    \begin{equation}
        \frac{1}{2} \Vec{x}^T \Vec{P} \Vec{x} + \Vec{q}^T\Vec{x}\mcm
        \label{eqn:quadprog}
    \end{equation}
    where $\Vec{P}$ and $\Vec{q}$ are a known matrix and vector respectively. One
    is allowed to formulate linear constraints for these unknowns in $\Vec{x}$,
    \begin{equation}
        \Vec{G}\Vec{x} \le \Vec{h} \mpt
        \label{eqn:linconstr}
    \end{equation}
    Here, $\Vec{G}$ and $\Vec{h}$ are again a known matrix and vector, and the
    inequality sign is to be interpreted as a component-wise inequality. Similarly,
	one is allowed to impose a linear equality constraint
    \begin{equation}
        \Vec{A}\Vec{x} = \Vec{b} \mcm
        \label{eqn:lineqconstr}
    \end{equation}
    as well as hard lower and upper bounds for the values in $\Vec{x}$.

    \begin{figure*}
        \centering
        \includegraphics[width=\textwidth]{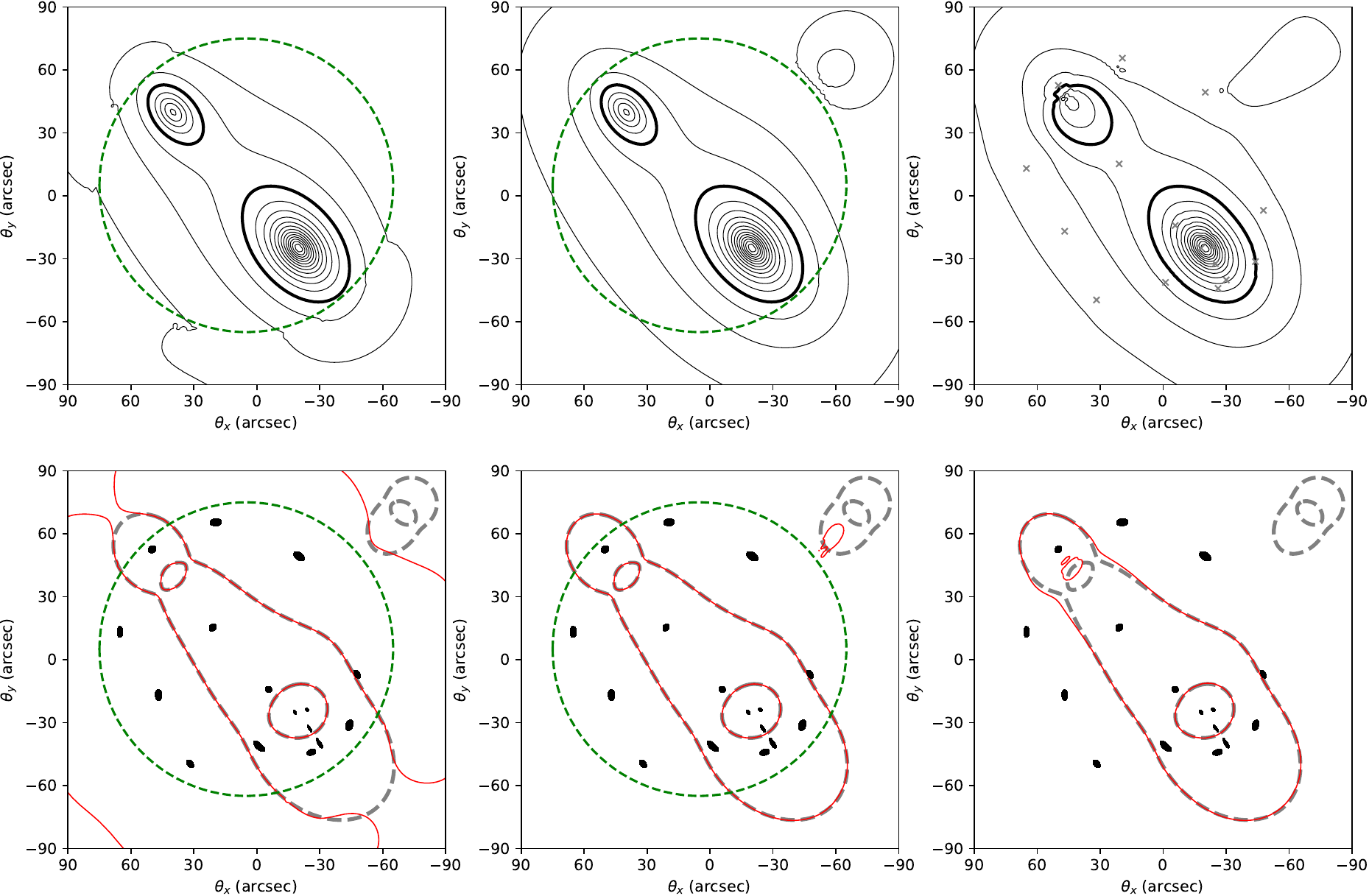}
        \caption{Results obtained when trying to remove the top-right density peak
        of the toy model, all using the {\sc mosek} solver. The top row shows the
        mass densities, the bottom row the re-calculated images using the adjusted
        model as well as the critical lines. For reference the original critical
        lines can be seen as thick dashed lines. In the left hand column, the 
        lens potential values inside the circular region were kept fixed, and
        allowed to change freely outward. In the center column, the same settings
        were used, only adding the additional constraint that the lens
        potential values at the border of the $\psi_{ij}$ grid should be preserved as well.
        In the right hand column,
        the border constraint was kept, but the one from the circular region
        was replaced by the regions of the images themselves.}
        \label{fig:simpeakremoved}
    \end{figure*}

    For simplicity, let us first consider a one dimensional example. Suppose the
    lensing potential is defined on a grid of $N$ points, $\psi_i$, $i=1,\ldots N$.
    Some of these values will be held fixed throughout the procedure, this will be
    denoted as $\psi_i = \tilde{\psi}_i$, while others are the values
    to be retrieved, described by $\psi_i = x_j$. There will be $M$ such unknown values
    $x_j$, $j=1,\ldots M$.

    Leading to a measure of the mass density through equation (\ref{eq:nablapsi}), the
    Laplacian will be of particular interest. For a discretized version of the lens
    potential values, this can be approximated by an $R$-point kernel 
    $L_k$, $k=1,\ldots R$, which through a convolution then yields an estimate of the local
    density. As an example, using e.g. the difference of differences
    \begin{equation}
        \left( \psi_3 - \psi_2 \right) - \left( \psi_2 - \psi_1 \right)
        = \psi_1 - 2 \psi_2 + \psi_3
    \end{equation}
    to approximate the Laplacian in a one-dimensional scenario, the three-point kernel $\Vec{L} = [1,-2,1]$
    could be used.

    Seeking smooth density distributions, it is actually the gradient of the density
    that is part of the minimization process. To optimize for this gradient in a discrete
    setting, one would compare such convolutions
    around neighbouring points, leading to the minimization of the following cost function:
    \begin{equation}
        \Sigma_i \left(\Sigma_{k=1}^R \psi_{i+k-1} L_k - \Sigma_{k=1}^R \psi_{i+k} L_k \right)^2 \mcm
    \end{equation}
    where for simplicity the bounds of the summation over $i$ are assumed to be such
    that the indices stay valid. One can easily regroup terms to yield
    \begin{equation}
        \Sigma_i \left(\Sigma_{k=1}^{R+1} \psi_{i+k-1} K_k\right)^2
        \label{eq:1doptproblem}
    \end{equation}
    for a now slightly larger kernel $K$. While this derivation was inspired by the gradient of the
    Laplacian, this general formula can be useful for many situations.
    For example, one could also use the Laplacian kernel directly to minimize for the
    density, a straightforward kernel $[-1, 1]$ to minimize the
    gradient of the lens potential, or even use multiple kernels with different weights.
    The latter would merely add more terms to the outer sum, optionally weighting them
    by a different factor, but the problem remains the same otherwise.

    This optimization problem is of the quadratic programming kind that was described
    above. To see this, let us focus on one particular term of the outer sum, say $i=7$. Assuming some
    of the $\psi$ values are fixed and some are to be optimized for, this term could
    look like the following:
    \begin{equation}
        (\psi_7 K_1 + \psi_8 K_2 + \psi_9 K_3)^2 = (\tilde{\psi_7} K_1 + x_1 K_2 + x_2 K_3)^2 \mpt
        \label{eq:1dsingleterm}
    \end{equation}
    Working out the square and leaving out the constant value that does not play a role
    during the optimization process, this can be written as
    \begin{equation}
        \begin{array}{r}
        \frac{1}{2}\left[ x_1 \; x_2\right] \left[\begin{array}{cc}
                                            2 K_2^2 & 2 K_2 K_3 \\
                                            2 K_2 K_3 & 2 K_3^2
                                               \end{array}\right]
                                         \left[\begin{array}{c}
                                             x_1 \\
                                             x_2
                                         \end{array}\right] \\
        + \left[ 2 \tilde{\psi}_7 K_1 K_2 \; 2 \tilde{\psi_7} K_1 K_3 \right] 
          \left[ \begin{array}{c}
                    x_1 \\
                    x_2
                 \end{array}\right] \mpt
        \end{array}
    \end{equation}
    For the full number of $M$ unknowns $x_j$, one can easily imagine the
    matrices containing the $K_k$ factors to be padded with zeros. To account for the 
    outer summation over $i$, the matrices for each of the terms simply need to be 
    added together. One then ends up with the optimization problem as formulated 
    in (\ref{eqn:quadprog}). Note that the resulting matrix $\Vec{P}$ will be a
    sparse matrix, in this one-dimensional example a so-called band matrix.

    There is also a constraint that needs to be taken into account: the mass
    density that is determined by the reconstructed potential, needs to be
    positive. This means that the convolution with the discrete Laplacian
    kernel $L_k$ should be positive for every grid point:
    \begin{equation}
        \Sigma_k \psi_{i+k-1} L_k \ge 0
    \end{equation}
    where $i$ is any value that leads to an actual constraint on some $x_j$
    values (i.e. not all $\psi_i$ values are fixed). The set of resulting
    constraints can easily be organized into the form of (\ref{eqn:linconstr}).
	Formulating additional constraints this way for the mass density can
	of course be done as well: perhaps one would like the mass in certain
	regions to lie within certain bounds, or even be equal to specific
	values. Similarly, one could add constraints for the gradient of
    $\psi_i$, i.e. the deflection angle, using the $[-1, 1]$ kernel.
    These desired properties would merely add rows to equations
	(\ref{eqn:linconstr}) or (\ref{eqn:lineqconstr}).

    For the two-dimensional case, of course a two-dimensional convolution will
    be needed, and the outer sum will need to be replaced by two sums, one for
    every grid dimension. This means that equation (\ref{eq:1doptproblem}) will
    be modified into the following:
    \begin{equation}
        \Sigma_i \Sigma_j \left(\Sigma_k\Sigma_l \psi_{i+k-1,j+l-1} K_{kl}\right)^2\mpt
    \end{equation}
    When writing out just a single term that is squared, it will have a comparable
    form as equation (\ref{eq:1dsingleterm}), implying that a very similar
    organization into a quadratic programming problem can take place. The structure
    of the $\Vec{P}$ matrix will no longer be that of a band matrix, but since the
    kernel is typically very small compared to the full grid size of the lens
    potential values, it will still be a sparse matrix.

    When using a kernel with e.g. values such as $[-1, 1]$, the result is only
    an estimate of the gradient up to some scale factor, depending on the grid
    resolution. Appendix~\ref{app:calib} describes a practical way in
    which this scale factor is determined in the code.

\section{Applications}\label{sec:apps}

    To numerically solve the quadratic programming problem, we made use of the 
    Python module {\sc qpsolvers} \citep{Caron_qpsolvers_Quadratic_Programming_2023}. This does not provide an
    implementation for the QP optimization by itself, but instead standardizes the formulation
    of the problem and still allows one to select one of several supported solver
    implementations, both open source and commercial. The actual solvers that were
    used in the examples below, are the {\sc mosek} \citep{mosek} software and
    the Splitting Conic Solver (SCS) \citep{scs,ocpb:16,odonoghue:21}.
    
    For the convolution kernel that is used as the discretized Laplace operator,
    various kernels with different extents can be used. In practice,
    the commonly used
    \begin{equation}
        \Vec{L} = \left[ 
            \begin{array}{ccc}
                0 & 1 & 0 \\ 
                1 & -4& 1 \\
                0 & 1 & 0  
            \end{array}
            \right]
        \label{eq:bbkernel}
    \end{equation}
    did not work as well as the following $5\times 5$ kernel from \citet{PrincImProc}:
    \begin{equation}
        \Vec{L} = \left[ 
            \begin{array}{ccccc}
                0 & 0 &  1 & 0 & 0 \\ 
                0 & 1 &  2 & 1 & 0 \\
                1 & 2 &-16 & 2 & 1 \\
                0 & 1 &  2 & 1 & 0 \\
                0 & 0 &  1 & 0 & 0 
            \end{array}
            \right] \mpt
        \label{eq:bbkernel}
    \end{equation}

    \subsection{Outside density peak in toy model}

        The left and center columns of Fig.~\ref{fig:simpeakremoved} show the results
        of the QP optimization when keeping the projected potenial values inside the
        circle fixed to those of the $128\times 128$ approximation of the toy model
        lens. For both situations the weights for the different kernel contributions
        were the same, but each has a different boundary constraint for the $\psi_{ij}$ grid.
        In the left situation,
        there was the requirement that the density that is calculated from the potential
        should equal the original one at the border. In the center situation the
        potential values themselves at the border were fixed in addition to those in
        the circular region.

        The additional freedom for the situation from the left hand panel allowed the
        solver ({\sc mosek} was used here) to erase the upper-right mass peak nearly
        completely. As the bottom part of the figure shows, the resulting critical
        line structure has become somewhat peculiar, clearly differing from the
        original one. This is much less the case for the center panel situation.
        The extra constraint on the border values of the lens potential did prevent
        the optimization to complete eliminate the top-right mass peak, although it
        is clearly reduced significantly.

        Note that because the lens potential
        values are unchanged within the circular region that contains the images,
        the image plane RMS is the same as in Table~\ref{tab:phiapproxprops} for the
        $128\times 128$ grid, on which these optimizations were based.

    \subsection{Monopole degeneracy revisited}

        While keeping lens potential values inside the circular region fixed will
        certainly preserve all properties in this strong lensing region, one
        could also keep only the potential values at the image locations fixed.
        That too will preserve all properties at those locations, but allows
        for changes in between the images as well. In this sense, it is quite
        similar to the monopole degeneracy from \citet{Liesenborgs4}, where
        specific basis functions were used to manipulate the mass density in
        between the images but without affecting the properties at the image
        locations themselves. The right hand column of Fig.~\ref{fig:simpeakremoved}
        shows the results for the toy model, where the settings were the same
        as in the center panel but the constraint of the circular region was 
        replaced by constraints at the image locations. As appendix~\ref{app:cl0024}
        shows, for CL0024+1654 the result is very similar to that of \citet{Liesenborgs4}.

\begin{figure}
            \centering
            \includegraphics[width=0.45\textwidth]{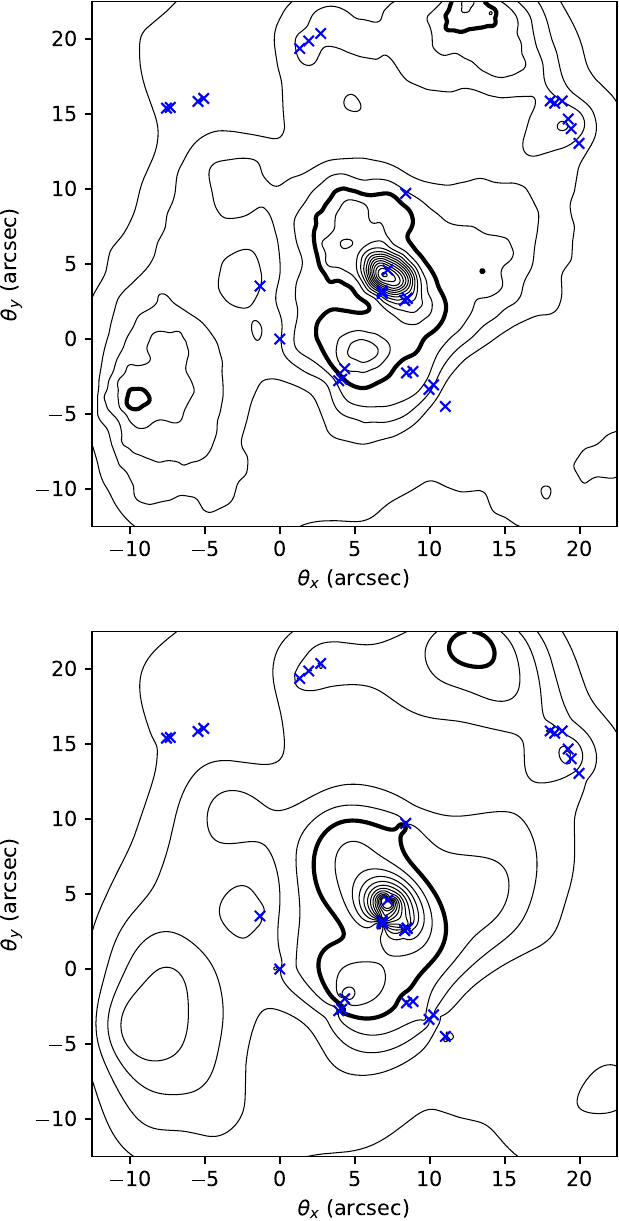}
            \caption{The upper panel shows the same model as the one in the top-left part of fig.~2 of
            \citet{2024MNRAS.527.2639P}. The feature around $(6,-1)$ arcsec, also called
            the south-east mass clump in the article, was an interesting result
            that was reproduced quite consistently in the inversions. The bottom
            panel shows how the extrapolation method described here, can cause
            mass to be redistributed thereby making this clump considerably less
            prominent. The optimization was performed with the SCS solver.}
            \label{fig:j1004redist}
        \end{figure}

        It is then interesting to apply this procedure to a mass model for SDSS~J1004+4112
        that was published recently \citep{2024MNRAS.527.2639P}, showing a peculiar mass
        concentration. For reference, the model is reproduced in the top panel of
        Fig.~\ref{fig:j1004redist}, where the mass density feature can be seen
        around $(6,-1)$ arcsec. The extrapolation procedure described above, can
        then be used to redistribute mass in between the image locations. Because
        the projected potential values in a small area around each image need to
        be conserved (to conserve all relevant derivatives), the resolution of the
        potential grid $\psi_{ij}$ affects how much freedom remains to redistribute
        the mass clump in question. Here, a very fine grid of $1536\times 1536$
        points was used, leading to the equivalent lens model shown in the bottom
        part of the figure. The procedure clearly was not as successful in erasing
        the mass clump as in the case of CL0024+1654, although the feature has become
        less prominent. That it is more difficult to remove it altogether could be
        expected due to the proximity of the images: since the projected potential is
        preserved in a small region around each image, so is the density and even
        its gradient. Note that in trying to erase certain features, the application
        of the procedure has also introduced some less desirable ones: a pinching
        effect can be seen e.g. around images (8,10) and (7,3).

    \subsection{Mass sheet degeneracy and source-position transformation}

        \begin{figure*}
            \centering
            \includegraphics[width=\textwidth]{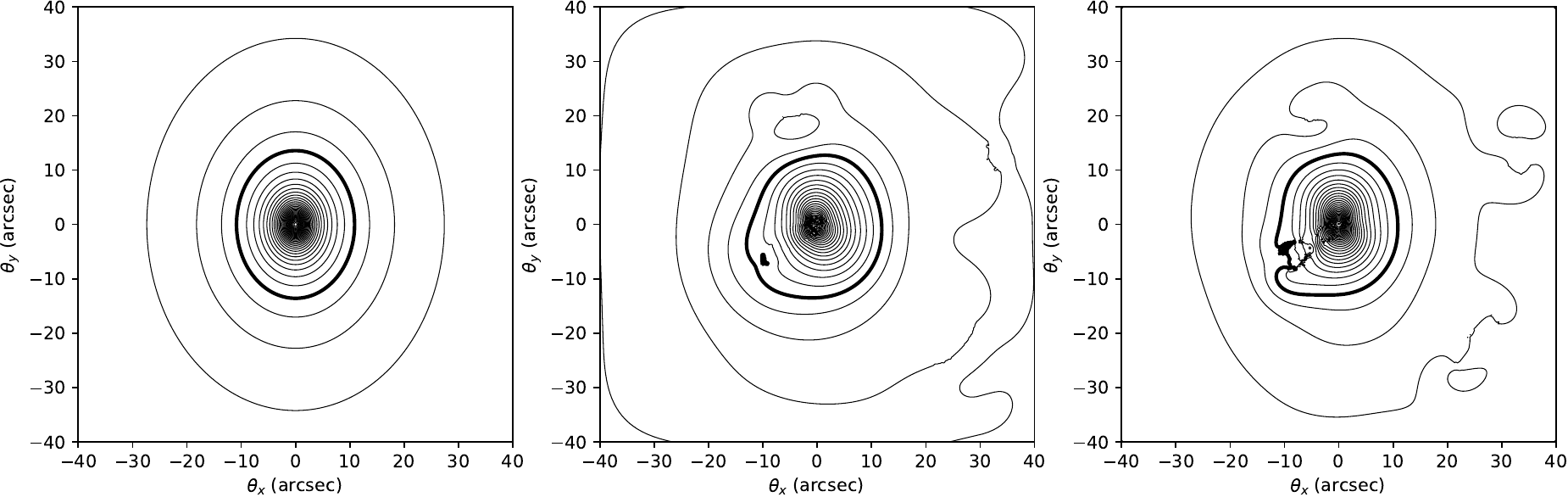}
            \includegraphics[width=\textwidth]{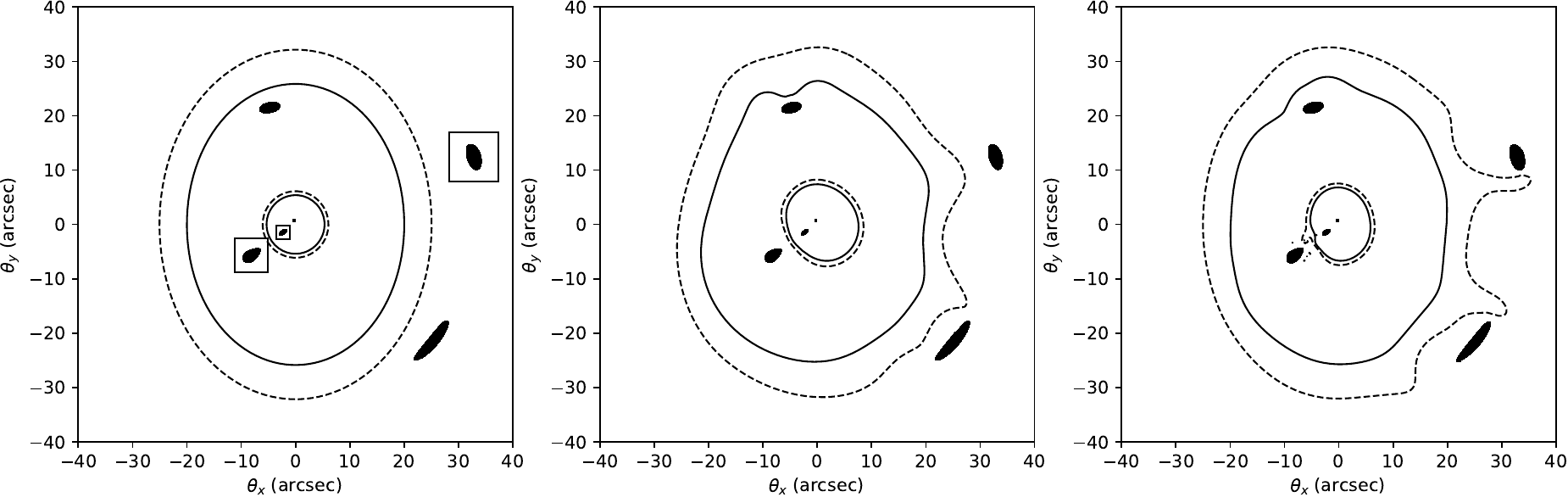}
            \includegraphics[width=\textwidth]{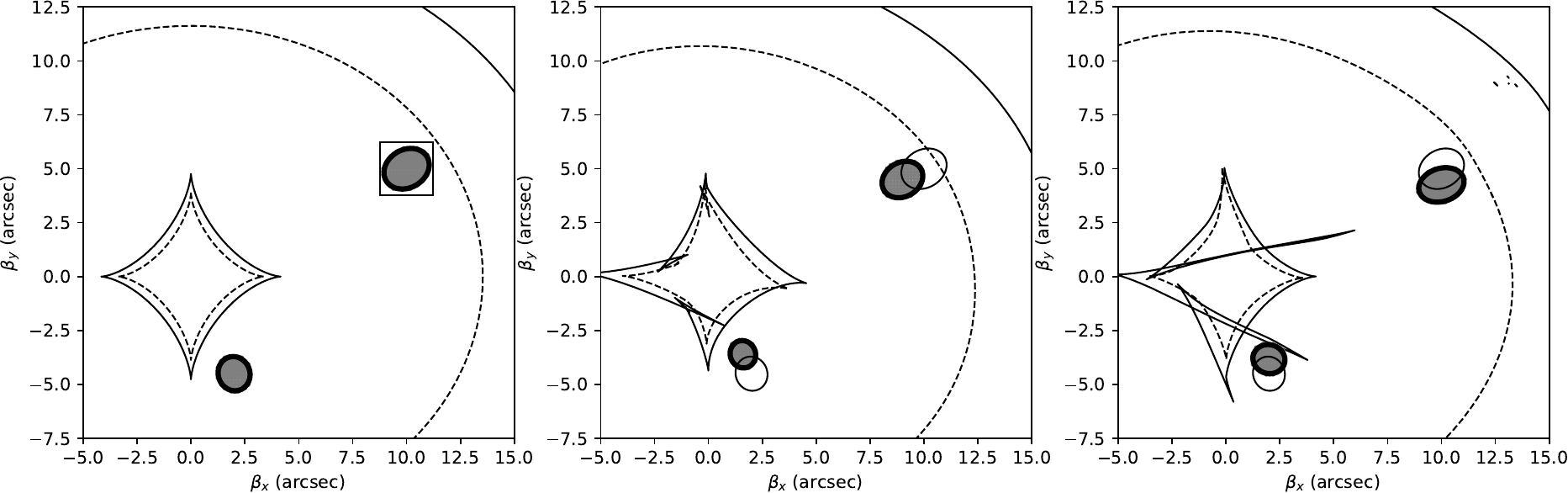}
            \caption{The left column shows the simulated lensing situation that is
                used to illustrate how the extrapolation procedure can be used
                to obtain generalized MSD and SPT variations
                of this lens. The top row shows the mass densities, where contours
                are again in units of $\kappa$ (for a source at $z=1.2$), spaced by 0.2, 
                and the thick like corresponds to $\kappa = 1$. The next row shows
                the images and critical lines that correspond to the sources and
                caustics in the bottom row. The source and images enclosed by square
                correspond to a redshift of 1.2, the other source is at a redshift
                of 1.8. The center column shows the model, image plane and source plane
                when a generalized MSD was constructed (see text). The orignal source
                positions are shown as thin lines. Similarly, the right column shows
                these properties for lens that differs by the SPT.}
            \label{fig:msdspt}
        \end{figure*}

        A relatively straightforward degeneracy is typically referred to as the
        mass-sheet degeneracy (MSD) \citep{FalcoMassSheet} or steepness degeneracy
        \citep{SahaSteepness}. Replacing the density $\Sigma(\Vec{\theta})$
        of a lens model by
        \begin{equation}
            \Sigma'(\Vec{\theta}) = \lambda \Sigma(\Vec{\theta}) - (1-\lambda)\Sigma_{\rm cr}
            \label{eq:masssheetsimple}
        \end{equation}
        yields a new model for which a source plane that is scaled in each dimension by the
        factor $\lambda$, corresponds to the same image plane. Note that this particular
        construction can be done for only a single source redshift, as the sheet of
        mass $\Sigma_{\rm cr}$ depends on the angular diameter distances to this source.
        This degeneracy does not preserve all properties though: since it scales the
        source plane, the magnifications of the images are changed by a factor of
        $\lambda^2$ (note that the relative magnifications are not affected). It can also 
        be shown that the time delays involved are scaled by this factor $\lambda$.

        One generalization of this degeneracy, using the extrapolation procedure
        from before, would be to find an alternative to the sheet itself: in the regions
        of the images, one could determine the projected potential values for the mass
        sheet $\Sigma_{\rm cr}$. Keeping these values fixed, the lens potential in other
        regions could be extrapolated, yielding a model $\Sigma_{\rm cr,eq}(\Vec{\theta})$ that has an equivalent effect
        as the mass sheet in this particular case. Using this in a similar way as in equation (\ref{eq:masssheetsimple})
        again produces a new $\Sigma'$ that has the same effect as the original mass
        sheet degeneracy. Since the extrapolation can be done in many ways, even for a
        single $\lambda$ value this leads to a multitude of equivalent mass models.

    \begin{figure*}
        \centering
        \includegraphics[width=0.95\textwidth]{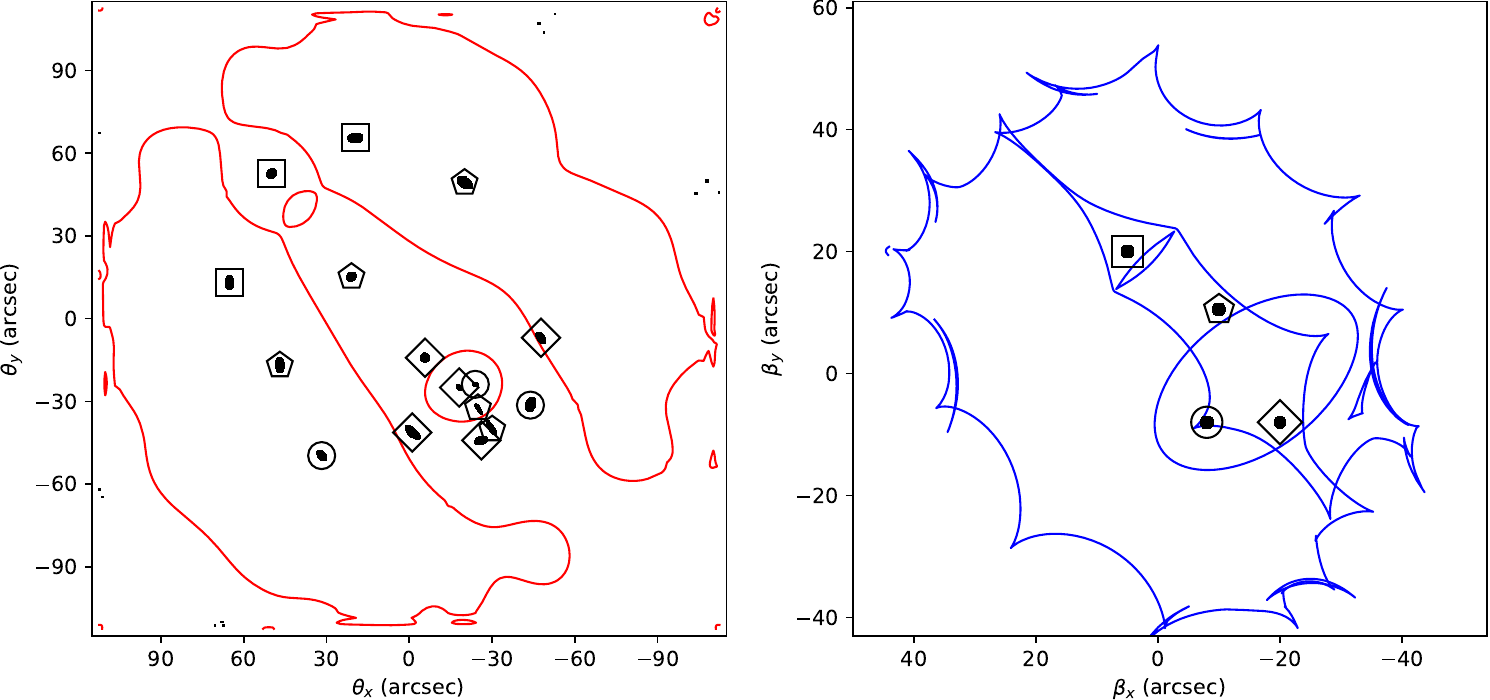}
        \caption{The left hand panel is a wider view of the peculiar critical lines
            that were shown in the left of Fig.~\ref{fig:simpeakremoved}, the right
            hand panel shows the corresponding caustics. Apart from the original
            multiple image systems, several other, small images are now predicted 
            near the border as well.}
        \label{fig:extralines}
    \end{figure*}

        It has been shown before that this degeneracy can be extended to multiple
        source redshifts, but with the same scale factor $\lambda$ \citep{Liesenborgs3},
        and even that different scale factors can be used \citep{2012MNRAS.425.1772L}.
        This more general scenario can also be inspected using the extrapolation
        procedure, although in this case the focus will be more on the constraints
        than on the potential values themselves. To illustrate this, consider the
        situation in the left hand column of Fig.~\ref{fig:msdspt}, which uses the
        same settings as in the example from \citet{2012MNRAS.425.1772L}: a
        non-singular isothermal ellipse (NSIE) mass distribution at a redshift of
        $0.5$, transforms the two elliptical sources at redshifts of $1.2$ and $1.8$
        from the bottom panel into the images of the center panel.

        Focusing temporarily on the first source only, a new model
        can be created using equation (\ref{eq:masssheetsimple}) and factor $\lambda_1$
        that causes a scaled version of the source to correspond to the same images.
        Let us identify the regions of the images as $\{\Vec{\theta}_{\rm reg,1}\}$,
        and save the deflections angles for this new model in these regions,
        $\Vec{\hat{\alpha}}'_1(\{\Vec{\theta}_{\rm reg,1}\})$, for later use. Of course,
        for the second source the same procedure can be performed, with a different
        $\lambda_2$, again producing deflection angles $\Vec{\hat{\alpha}}'_2(\{\Vec{\theta}_{\rm reg,2}\})$,
        now in the regions of the other images.

        The goal is now to create a single lens model that has both $\Vec{\hat{\alpha}}'_1(\{\Vec{\theta}_{\rm reg,1}\})$
        and $\Vec{\hat{\alpha}}'_2(\{\Vec{\theta}_{\rm reg,2}\})$ as deflection angles,
        and it is this which the extrapolation method can produce. There will be
        no\footnote{Actually, in practice a single value is fixed on the $\psi_{\rm i,j}$ grid,
        so that the procedure has some value to start from. Since the offset of the lens
        potential is irrelevant, the precise value does not play a role.}
        values of the lens potential that are fixed, all will be calculated. Apart
        from the constraint that keeps the mass density positive, there are now also
        constraints for the gradients of the projected potential, as these correspond
        to the deflection angles $\Vec{\hat{\alpha}}'$ of the new model. For such gradients
        the kernel $[-1, 1]$ can be used in $x$- and $y$-directions, yielding estimates
        of $\Vec{\hat{\alpha}}'$. In principle, equation (\ref{eqn:lineqconstr})
        could enforce these values in the regions of the images to be respectively
        $\Vec{\hat{\alpha}}'_1(\{\Vec{\theta}_{\rm reg,1}\})$ and $\Vec{\hat{\alpha}}'_2(\{\Vec{\theta}_{\rm reg,2}\})$,
        but we found that this exact constraint does not work well in practice. Instead,
        some small deviations are allowed, which can be formulated as inequality constraints
        using equation (\ref{eqn:linconstr}).
        The center column of Fig.~\ref{fig:msdspt} shows a model that was obtained using
        this procedure, for $\lambda_1 = 0.9$ and $\lambda_2 = 0.8$. The sources that
        were scaled by these factors are shown in the bottom panel, and cause the same
        images as before to be generated, as can be seen in the center panel.

        The mass-sheet degeneracy is a special case of the source-position transformation (SPT)
        described in \citet{2014A&A...564A.103S}. In this more general version, the
        source plane does not undergo a mere rescaling, but can be transformed in a more
        general way. In case one would scale only the $y$-dimension of the source plane
        by a factor $\lambda$, one can write
        \begin{equation}
            \beta'_y = \lambda \beta_y = \theta_y - \frac{D_{\rm ds}}{D_{\rm s}} \hat{\alpha}'_y \mcm
        \end{equation} 
        where
        \begin{equation}
            \hat{\alpha}'_y = \frac{D_{\rm s}}{D_{\rm ds}}\theta_y(1-\lambda) + \lambda \hat{\alpha}_y \mpt
        \end{equation}
        Based on the original deflection angles in the image regions, this way one
        can calculate the constraints for
        $\Vec{\hat{\alpha}}'_1(\{\Vec{\theta}_{\rm reg,1}\})$ and $\Vec{\hat{\alpha}}'_2(\{\Vec{\theta}_{\rm reg,2}\})$,
        which can subsequently be solved in the same way as before. The right hand
        column of Fig.~\ref{fig:msdspt} shows how the two source shapes that were
        rescaled in the $y$-direction by a factor $\lambda=0.85$ produce the same
        images again. In practice, obtaining results that correspond to this type
        of SPT turned out to be quite difficult. In particular, when further decreasing $\lambda$
        one quickly needed to allow more and more deviations in the targeted $\Vec{\hat{\alpha}}'$
        values for the QP procedure to still find a solution. Similar difficulties
        were also noted in \citet{2014A&A...564A.103S}, as more arbitrary changes of
        the source plane are no longer guaranteed to be compatible with deflection
        angles being the gradient of the lensing potential.

        As with the MSD itself, while these variations still generate (nearly) the
        same images, other properties are no longer preserved. The time delays will
        change in a less predictable way, and magnifications themselves are changed
        as well.

\section{Discussion and conclusion}\label{sec:discussion}

    In this article we have looked at the problem of lensing degeneracies from the
    perspective of the projected potential. Starting from a working lens model and keeping
    the lens potential values fixed in the relevant regions, i.e. at least the regions
    of the images in the system, one can obtain a multitude of equivalent lens models
    that preserve all lensing properties. If one concentrates on only retaining the
    deflection angles, models compatible with the same observed images can still be
    obtained, however these will no longer necessarily preserve other properties.
    This also illustrates the different constraining power of different types of
    observations. Time delay measurements help probe the projected potential directly
    and are therefore of particular importance. Observed images provide information
    about the gradient of the potential, which illustrates why, without fixing the
    underlying shape of the lens model, many multiple images systems are required
    to build the lens potential from its gradients. Weak lensing measurement probe
    the curvature of the potential, providing the least detailed information.

    A method using quadratic programming was described with which such degeneracies
    can be explored in practice. Different constraints or weights of the convolution
    kernels yield different results, and even different solvers with otherwise the
    same settings typically do no converge to the exact same solution. All solutions
    do preserve the desired lensing properties, further indicating that multiple
    solutions can explain the same observations.

    \begin{figure*}
        \centering
        \includegraphics[width=\textwidth]{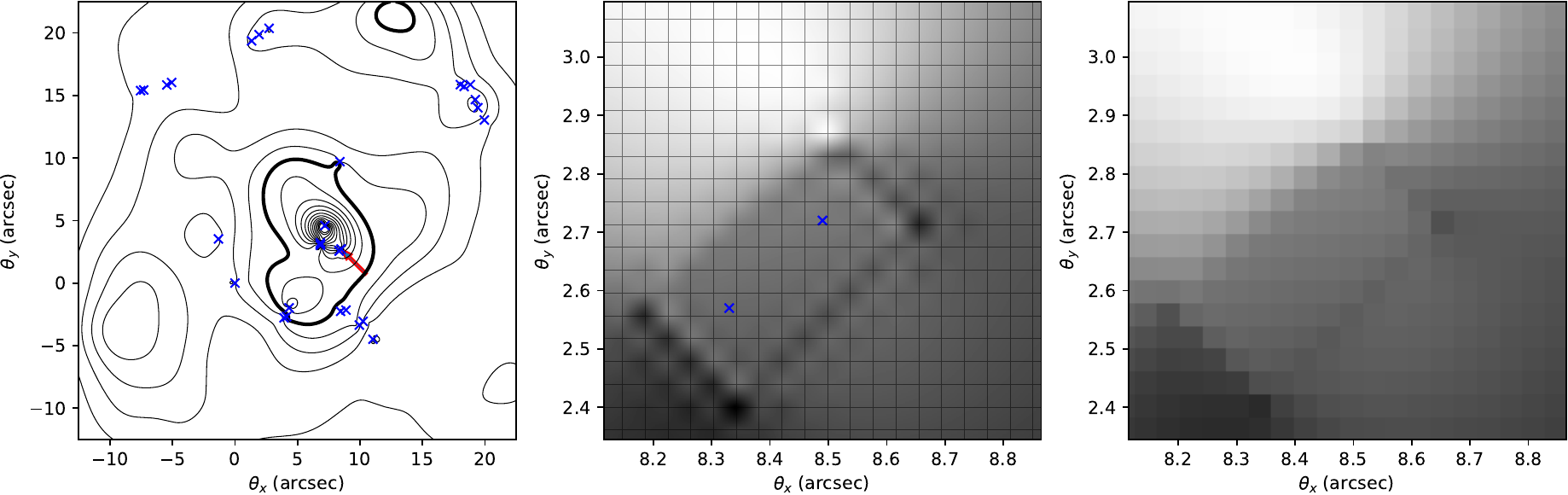}
        \caption{For the SDSS J1004+4112 images that are indicated by the arrow in the left panel,
            the center panel shows the density in their neighbourhood. The locations
            of the potential grid values are the intersections of horizontal and
            vertical lines. The right hand panel shows what the QP procedure calculates
            for the density, based on the convolution of these same potential values 
            with the Laplacian kernel from equation~(\ref{eq:bbkernel}).
            Comparing the two shows that the small scale fluctuations in the center
            panel can at least partly be explained by a mismatch between the different
            ways the density is calculated.}
        \label{fig:densmismatch}
    \end{figure*}

    At least using this particular method, it is
    is not always easy to eliminate the light-free mass features that \grale{} reconstructs.
    They can be partially smoothed by redistributing mass over the lens plane, but 
    they do not necessarily go away completely. The work presented here shows that 
    because of lensing degeneracies, one should not take the specific shapes of such
    mass features too seriously, but on the other hand one should not merely dismiss them 
    as artifacts of a free-form reconstruction. This also lends support for the
    reality of the dark matter clumps in SDSS~J1004+4112 as well as in Abell~1689
    \citep{2023MNRAS.525.2519G}, which could be similar, but somewhat less massive 
    than the ones found in the Coma cluster using weak lensing \citep{2010ApJ...713..291O}.

    The left hand panel of Fig.~\ref{fig:simpeakremoved} showed how certain settings
    could remove the mass peak from the toy model completely. The price was the
    rather odd critical line structure, for which a more wide field view can be
    seen in the left panel of Fig.~\ref{fig:extralines}. The area shown borders the region on which
    the lens potential is defined, explaining the behaviour of the lines near the
    boundaries. The right hand panel of the same figure shows the corresponding
    caustic structure. While this particular lens model still generates the same
    images as the true lens, several extra, smaller images are predicted as well.
    Unfortunately this is not something that can be prevented directly using the quadratic
    programming method.
	These strange critical line and caustic structures identify a possible disadvantage of thinking in terms
	of the lens potential. While this does allow for a large amount of flexibility,
	it does not provide an overview of the mass distribution in its entirety. Similar
	to the notion of the external shear, it can easily encode lensing effects without
	having to identify any part of the mass density as the origin of these effects.
	In this respect, explicitly modelling the mass density itself has the advantage,
	as there simply is no other mass needed that is not part of the model.

    Not always visible in the figures as they are shown here, are some rather unphysical
    fluctuations in the density near the images where the potential values are
    retained. Fig.~\ref{fig:densmismatch} illustrates this effect: for the image
    in the SDSS J1004+4112 system that is indicated in the left panel, the center
    panel shows the density according to the model that is based on the extrapolated
    lens potential values. The intersections of the horizontal and vertical lines
    identify the grid where these potential values are defined. To conserve the lens
    properties near the images, some of these values were kept fixed; these result
    in the density region that resembles a rotated rectangle. Adjacent to this region,
    where the extrapolation starts, is where some fluctuations in the density can be
    seen.

    This seems to be explainable, at least in part, to a slight mismatch between the
    final model using bicubic interpolation of the grid values and the optimization
    procedure that estimates the density using a convolution kernel. The first one
    is used to have a lens model for which all properties can be calculated at any
    point, even between the grid points. The second is needed to be able to formulate
    the optimization as a QP problem. The right hand panel shows the estimated density from the
    convolution, implying that this is the situation that is considered by the QP
    procedure. While the densities are no longer defined everywhere, but only on the
    grid points, the result does seem to suffer much less from the fluctucations from
    the center panel. Even so, the area where the potential values are preserved can
    still be identified clearly, meaning that the extrapolation result is not as good
    as one would hope.

    It is not yet clear how this effect should best be addressed. Perhaps a convolution
    kernel that is better suited than the one from equation~(\ref{eq:bbkernel}) can
    be found, although initial attempts have not been successful. Another approach could
    be to no longer use the QP formulation of the problem, but another optimization
    strategy that does not require the approximation with the convolution kernel. Possibly
    such an alternative optimization procudure could prevent the prediction of 
    unobserved images as well.

\section*{Data availability}

    The data underlying this article will be shared on reasonable request
    to the corresponding author.

\section*{Acknowledgments}
    
    JL acknowledges the use of the computational resources and services 
    provided by the VSC (Flemish Supercomputer Center), funded by the 
    Research Foundation - Flanders (FWO) and the Flemish Government.

\bibliographystyle{mnras}
%\bibliography{paper}

\begin{thebibliography}{}
\makeatletter
\relax
\def\mn@urlcharsother{\let\do\@makeother \do\$\do\&\do\#\do\^\do\_\do\%\do\~}
\def\mn@doi{\begingroup\mn@urlcharsother \@ifnextchar [ {\mn@doi@}
  {\mn@doi@[]}}
\def\mn@doi@[#1]#2{\def\@tempa{#1}\ifx\@tempa\@empty \href
  {http://dx.doi.org/#2} {doi:#2}\else \href {http://dx.doi.org/#2} {#1}\fi
  \endgroup}
\def\mn@eprint#1#2{\mn@eprint@#1:#2::\@nil}
\def\mn@eprint@arXiv#1{\href {http://arxiv.org/abs/#1} {{\tt arXiv:#1}}}
\def\mn@eprint@dblp#1{\href {http://dblp.uni-trier.de/rec/bibtex/#1.xml}
  {dblp:#1}}
\def\mn@eprint@#1:#2:#3:#4\@nil{\def\@tempa {#1}\def\@tempb {#2}\def\@tempc
  {#3}\ifx \@tempc \@empty \let \@tempc \@tempb \let \@tempb \@tempa \fi \ifx
  \@tempb \@empty \def\@tempb {arXiv}\fi \@ifundefined
  {mn@eprint@\@tempb}{\@tempb:\@tempc}{\expandafter \expandafter \csname
  mn@eprint@\@tempb\endcsname \expandafter{\@tempc}}}

\bibitem[\protect\citeauthoryear{Burger \& Burge}{Burger \&
  Burge}{2009}]{PrincImProc}
Burger W.,  Burge M.~J.,  2009, Principles of Digital Image Processing:
  Fundamental Techniques, 1 edn.
Springer Publishing Company, Incorporated

\bibitem[\protect\citeauthoryear{Caron et~al.,}{Caron
  et~al.}{2023}]{Caron_qpsolvers_Quadratic_Programming_2023}
Caron S.,  et~al., 2023, {qpsolvers: Quadratic Programming Solvers in Python,
  version 3.5.0}, \url {https://github.com/qpsolvers/qpsolvers}

\bibitem[\protect\citeauthoryear{{Coles}}{{Coles}}{2008}]{2008ApJ...679...17C}
{Coles} J.,  2008, \mn@doi [\apj] {10.1086/587635}, \href
  {http://adsabs.harvard.edu/abs/2008ApJ...679...17C} {679, 17}

\bibitem[\protect\citeauthoryear{{Falco}, {Gorenstein}  \& {Shapiro}}{{Falco}
  et~al.}{1985}]{FalcoMassSheet}
{Falco} E.~E.,  {Gorenstein} M.~V.,   {Shapiro} I.~I.,  1985, \mn@doi [\apjl]
  {10.1086/184422}, \href {http://adsabs.harvard.edu/abs/1985ApJ...289L...1F}
  {289, L1}

\bibitem[\protect\citeauthoryear{{Ghosh} et~al.,}{{Ghosh}
  et~al.}{2021}]{2021MNRAS.506.6144G}
{Ghosh} A.,  et~al., 2021, \mn@doi [\mnras] {10.1093/mnras/stab1196}, \href
  {https://ui.adsabs.harvard.edu/abs/2021MNRAS.506.6144G} {506, 6144}

\bibitem[\protect\citeauthoryear{{Ghosh}, {Adams}, {Williams}, {Liesenborgs},
  {Alavi}  \& {Scarlata}}{{Ghosh} et~al.}{2023}]{2023MNRAS.525.2519G}
{Ghosh} A.,  {Adams} D.,  {Williams} L. L.~R.,  {Liesenborgs} J.,  {Alavi} A.,
   {Scarlata} C.,  2023, \mn@doi [\mnras] {10.1093/mnras/stad2418}, \href
  {https://ui.adsabs.harvard.edu/abs/2023MNRAS.525.2519G} {525, 2519}

\bibitem[\protect\citeauthoryear{{Jullo}, {Kneib}, {Limousin},
  {El{\'{\i}}asd{\'o}ttir}, {Marshall}  \& {Verdugo}}{{Jullo}
  et~al.}{2007}]{2007NJPh....9..447J}
{Jullo} E.,  {Kneib} J.,  {Limousin} M.,  {El{\'{\i}}asd{\'o}ttir} {\'A}.,
  {Marshall} P.~J.,   {Verdugo} T.,  2007, \mn@doi [New Journal of Physics]
  {10.1088/1367-2630/9/12/447}, \href
  {http://adsabs.harvard.edu/abs/2007NJPh....9..447J} {9, 447}

\bibitem[\protect\citeauthoryear{{Liesenborgs} \& {De Rijcke}}{{Liesenborgs} \&
  {De Rijcke}}{2012}]{2012MNRAS.425.1772L}
{Liesenborgs} J.,  {De Rijcke} S.,  2012, \mn@doi [\mnras]
  {10.1111/j.1365-2966.2012.21751.x}, \href
  {https://ui.adsabs.harvard.edu/abs/2012MNRAS.425.1772L} {425, 1772}

\bibitem[\protect\citeauthoryear{{Liesenborgs}, {De Rijcke}  \&
  {Dejonghe}}{{Liesenborgs} et~al.}{2006}]{Liesenborgs}
{Liesenborgs} J.,  {De Rijcke} S.,   {Dejonghe} H.,  2006, \mn@doi [\mnras]
  {10.1111/j.1365-2966.2006.10040.x}, \href
  {http://adsabs.harvard.edu/cgi-bin/nph-bib_query?bibcode=2006MNRAS.367.1209L&db_key=AST}
  {367, 1209}

\bibitem[\protect\citeauthoryear{{Liesenborgs}, {De Rijcke}, {Dejonghe}  \&
  {Bekaert}}{{Liesenborgs} et~al.}{2008a}]{Liesenborgs3}
{Liesenborgs} J.,  {De Rijcke} S.,  {Dejonghe} H.,   {Bekaert} P.,  2008a,
  \mn@doi [\mnras] {10.1111/j.1365-2966.2008.13026.x}, \href
  {http://adsabs.harvard.edu/abs/2008MNRAS.386..307L} {386, 307}

\bibitem[\protect\citeauthoryear{{Liesenborgs}, {De Rijcke}, {Dejonghe}  \&
  {Bekaert}}{{Liesenborgs} et~al.}{2008b}]{Liesenborgs4}
{Liesenborgs} J.,  {De Rijcke} S.,  {Dejonghe} H.,   {Bekaert} P.,  2008b,
  \mn@doi [\mnras] {10.1111/j.1365-2966.2008.13586.x}, \href
  {http://adsabs.harvard.edu/abs/2008MNRAS.389..415L} {389, 415}

\bibitem[\protect\citeauthoryear{{Liesenborgs}, {Williams}, {Wagner}  \& {De
  Rijcke}}{{Liesenborgs} et~al.}{2020}]{2020MNRAS.494.3253L}
{Liesenborgs} J.,  {Williams} L. L.~R.,  {Wagner} J.,   {De Rijcke} S.,  2020,
  \mn@doi [\mnras] {10.1093/mnras/staa842}, \href
  {https://ui.adsabs.harvard.edu/abs/2020MNRAS.494.3253L} {494, 3253}

\bibitem[\protect\citeauthoryear{{Meneghetti} et~al.,}{{Meneghetti}
  et~al.}{2017}]{2017MNRAS.472.3177M}
{Meneghetti} M.,  et~al., 2017, \mn@doi [\mnras] {10.1093/mnras/stx2064}, \href
  {https://ui.adsabs.harvard.edu/abs/2017MNRAS.472.3177M} {472, 3177}

\bibitem[\protect\citeauthoryear{O'Donoghue}{O'Donoghue}{2021}]{odonoghue:21}
O'Donoghue B.,  2021, {SIAM} Journal on Optimization, 31, 1999

\bibitem[\protect\citeauthoryear{O'Donoghue, Chu, Parikh  \& Boyd}{O'Donoghue
  et~al.}{2016}]{ocpb:16}
O'Donoghue B.,  Chu E.,  Parikh N.,   Boyd S.,  2016, Journal of Optimization
  Theory and Applications, 169, 1042

\bibitem[\protect\citeauthoryear{O'Donoghue, Chu, Parikh  \& Boyd}{O'Donoghue
  et~al.}{2022}]{scs}
O'Donoghue B.,  Chu E.,  Parikh N.,   Boyd S.,  2022, {SCS}: Splitting Conic
  Solver, version 3.2.3, \url{https://github.com/cvxgrp/scs}

\bibitem[\protect\citeauthoryear{{Okabe}, {Okura}  \& {Futamase}}{{Okabe}
  et~al.}{2010}]{2010ApJ...713..291O}
{Okabe} N.,  {Okura} Y.,   {Futamase} T.,  2010, \mn@doi [\apj]
  {10.1088/0004-637X/713/1/291}, \href
  {https://ui.adsabs.harvard.edu/abs/2010ApJ...713..291O} {713, 291}

\bibitem[\protect\citeauthoryear{{Perera}, {Williams}, {Liesenborgs}, {Ghosh}
  \& {Saha}}{{Perera} et~al.}{2024}]{2024MNRAS.527.2639P}
{Perera} D.,  {Williams} L. L.~R.,  {Liesenborgs} J.,  {Ghosh} A.,   {Saha} P.,
   2024, \mn@doi [\mnras] {10.1093/mnras/stad3366}, \href
  {https://ui.adsabs.harvard.edu/abs/2024MNRAS.527.2639P} {527, 2639}

\bibitem[\protect\citeauthoryear{{Saha} \& {Williams}}{{Saha} \&
  {Williams}}{2004}]{2004AJ....127.2604S}
{Saha} P.,  {Williams} L.~L.~R.,  2004, \mn@doi [\aj] {10.1086/383544}, \href
  {http://adsabs.harvard.edu/abs/2004AJ....127.2604S} {127, 2604}

\bibitem[\protect\citeauthoryear{{Saha} \& {Williams}}{{Saha} \&
  {Williams}}{2006}]{SahaSteepness}
{Saha} P.,  {Williams} L.~L.~R.,  2006, \mn@doi [\apj] {10.1086/508798}, \href
  {http://adsabs.harvard.edu/abs/2006ApJ...653..936S} {653, 936}

\bibitem[\protect\citeauthoryear{{Schneider} \& {Sluse}}{{Schneider} \&
  {Sluse}}{2014}]{2014A&A...564A.103S}
{Schneider} P.,  {Sluse} D.,  2014, \mn@doi [\aap]
  {10.1051/0004-6361/201322106}, \href
  {https://ui.adsabs.harvard.edu/abs/2014A&A...564A.103S} {564, A103}

\bibitem[\protect\citeauthoryear{{Schneider}, {Ehlers}  \& {Falco}}{{Schneider}
  et~al.}{1992}]{SchneiderBook}
{Schneider} P.,  {Ehlers} J.,   {Falco} E.~E.,  1992, {Gravitational Lenses}.
Springer-Verlag

\bibitem[\protect\citeauthoryear{{Sendra}, {Diego}, {Broadhurst}  \&
  {Lazkoz}}{{Sendra} et~al.}{2014}]{2014MNRAS.437.2642S}
{Sendra} I.,  {Diego} J.~M.,  {Broadhurst} T.,   {Lazkoz} R.,  2014, \mn@doi
  [\mnras] {10.1093/mnras/stt2076}, \href
  {https://ui.adsabs.harvard.edu/abs/2014MNRAS.437.2642S} {437, 2642}

\bibitem[\protect\citeauthoryear{{Torres-Ballesteros} \&
  {Casta{\~n}eda}}{{Torres-Ballesteros} \&
  {Casta{\~n}eda}}{2023}]{2023MNRAS.518.4494T}
{Torres-Ballesteros} D.~A.,  {Casta{\~n}eda} L.,  2023, \mn@doi [\mnras]
  {10.1093/mnras/stac3253}, \href
  {https://ui.adsabs.harvard.edu/abs/2023MNRAS.518.4494T} {518, 4494}

\bibitem[\protect\citeauthoryear{$\textrm{MOSEK ApS}$}{$\textrm{MOSEK
  ApS}$}{2023}]{mosek}
$\textrm{MOSEK ApS}$ 2023, MOSEK Optimizer API for Python 10.1.11.
\url {https://docs.mosek.com/latest/pythonapi/index.html}

\makeatother
\end{thebibliography}

\appendix
\section{Kernel scale factor calibration}\label{app:calib}

    To obtain the scale factor to use for a correct estimation of the
    gradient of the $\psi_{ij}$ grid, which corresponds to the
    deflection angles, one can use the fact that for a Singular
    Isothermal Sphere (SIS) lens model, the deflection angle is constant
    in size
    \begin{equation}
        \hat{\alpha} = 4\pi\sigma_v^2/c^2\mcm
        \label{eq:sisalpha}
    \end{equation}
    where $\sigma_v$ is the velocity dispersion for the model. Sampling the lens
    potential values of this model on a grid with a specific resolution, 
    convolving these with the gradient kernel and comparing the result to
    the value expected from equation (\ref{eq:sisalpha}), then yields the
    scale factor.

    A similar approach can be used to get the required scale factor
    when using the Laplacian kernel, e.g. equation~(\ref{eq:bbkernel}).
    In this case, a model can be constructed that corresponds to a sheet
    of a specific constant density $\Sigma_{\rm s}$. Again, obtaining the projected potential
    values on a grid with a certain resolution, convolving the values with
    the Laplacian kernel and comparing the result to the desired $\Sigma_{\rm s}$
    reveals the required scale factor for the kernel.

\section{Comparision with CL0024+1654 results}\label{app:cl0024}

\begin{figure*}
         \centering
         \includegraphics[width=\textwidth]{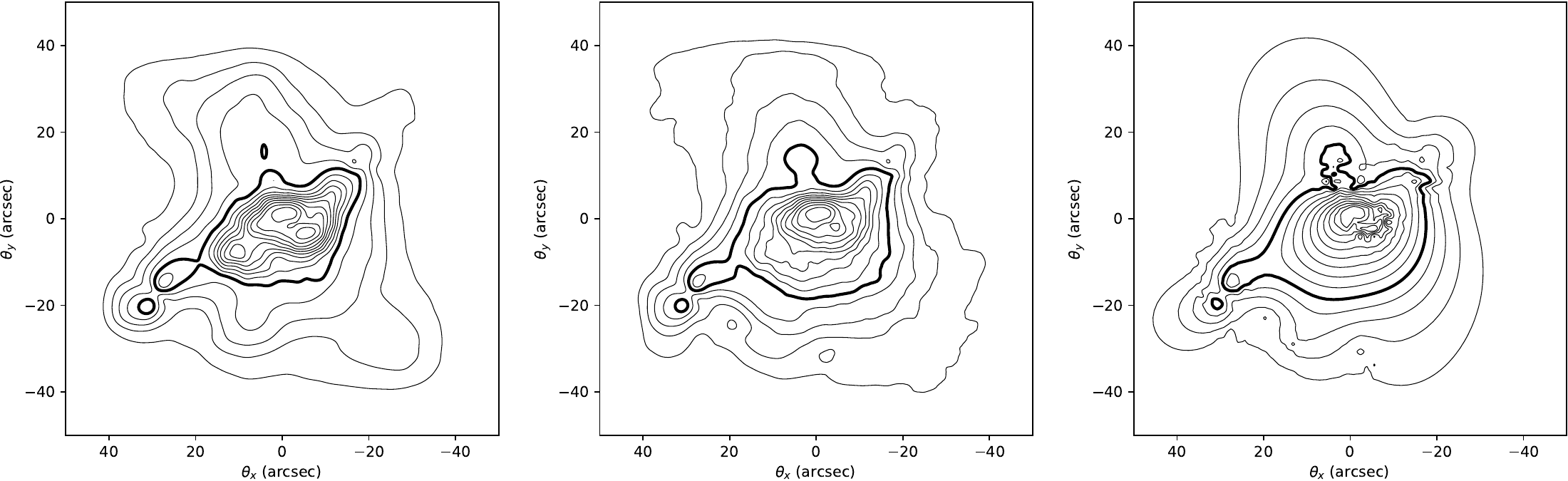}
         \caption{Applying the method from this article to the model from
            \citet{Liesenborgs4} (left panel), yields the result shown in the right
            panel. For comparison, the center panel shows the result that was obtained
            in the aforementioned article using the monopole degeneracy.}
         \label{fig:cl0024revis}
    \end{figure*}

    Fig.~\ref{fig:cl0024revis} compares the results from \citet{Liesenborgs4} to the ones
    obtained with the lens potential extrapolation method from this article. The left
    hand panel shows the lens inversion results that were obtained for CL0024+1654,
    were a small peak can be seen around $(10,-10)$ arcsec. As this was relatively far
    from the image locations, the monopole degeneracy was used to redistribute mass,
    obtaining the results from the center panel. The right hand panel then shows the
    results with the method from this article, where the {\sc mosek} solver was used.
    The resulting mass map shows an interesting correspondence to the results from the
    center panel where mass was redistributed explicitly.

\bsp 
\label{lastpage}

\end{document}